\documentstyle[12pt]{article}
\newcommand{\trup}{\bigtriangleup}
\newcommand{\bx}{{\bf x}}
\newcommand{\be}{\begin{equation}}
\newcommand{\ee}{\end{equation}}
\newcommand{\Res}{{\rm Res}}

\begin{document}

\begin{center}
{\large\bf Restrictions on negative energy density in a curved 
spacetime  }
\end{center}
\bigskip
\bigskip
\begin{center}
Dae-Yup Song\footnote[2]{Electronic address: 
        dsong@sunchon.sunchon.ac.kr}
\end{center}

\bigskip
\begin{center}
{\it Department of Physics,\\ Sunchon National University, Sunchon 
540-742, Korea}
\end{center}
\bigskip 

\begin{abstract}
Recently a restriction ("quantum inequality-type relation") on the 
(renormalized) energy density measured by a static observer in a 
"globally static" (ultrastatic) spacetime has been formulated by 
Pfenning and Ford for the minimally coupled scalar field, in 
the extension of quantum inequality-type relation on flat spacetime
of Ford and Roman. They found negative lower bounds for the line 
integrals of energy density multiplied by a sampling (weighting) 
function, and explicitly evaluate them for some specific spacetimes. 
In this paper, we study the lower bound on spacetimes whose spacelike 
hypersurfaces are compact and without boundary. 
In the short "sampling time" limit, the bound has asymptotic expansion.
Although the expansion can not be represented by locally invariant 
quantities in general due to the nonlocal nature of the integral, 
we explicitly evaluate the dominant terms in the limit in terms of the
invariant quantities. We also make an estimate for the bound 
in the long sampling time limit.
\end{abstract}
\begin{center}
( February 1997 )
\end{center}

\newpage

\section{Introduction}
Since spacetime curvature is produced by the total stress-energy
tensor of all the matter that inhabits spacetime, it is very 
important to investigate the conditions that stress-energy tensor
$T_{\mu\nu}$ would satisfy. Indeed, "energy conditions" i.e. 
assumptions 
concerning the positivity  of the locally measured energy density, 
play a key role in the proof of the classical theorems on the 
large-scale structure of spacetime \cite{HE,Wald}. 
In quantum field theory, however, it well known that 
the local (i.e. pointwise) energy conditions can be violated
for the expectation value of stress-energy tensor, as 
can be seen in the example of a free scalar field in Minkowski 
spacetime  when its state is the vacuum plus a small admixture of 
two-particle state \cite{FD}.

Instead of the local energy conditions, various nonlocal energy
condition which may be still relevant with the theorems on large-scale
structure are proposed. One of the most prominent conditions is the   
averaged null energy condition (ANEC) \cite{Tipler,Roman}
which requires 
\begin{equation}
\int_\gamma <T_{\mu\nu}> k^\mu k^\nu d\tau \geq 0,
\end{equation}
where the $\gamma$ is complete null geodesic, $k^\mu$ is the tangent 
vector along $\gamma$, and $\tau$ is an affine parameter. 
$<T_{\mu\nu}>$ is the (renormalized) expectation value 
evaluated on a state and, for the ANEC, the inequality is required 
to be satisfied for any state \cite{Klink}.
If the $\gamma$ is timelike and $\tau$ is the observer's proper time,
the requirement of Eq.(1) is the average weak energy condition (AWEC).
It has been known, however, that \cite{WY,Visser1}
the ANEC can be violated in general due to the anomalous scaling of 
renormalized expectation value of energy momentum tensor under scaling 
transformation, and explicit examples of violations of ANEC
have been given by Visser \cite{Visser2,Visser3} \cite{comment}. 
Some generalizations of ANEC has been proposed;
Ford and Roman evaluated the ANEC integrals (the left
hand side of Eq.(1))
with a weighting ("sampling") function
$f(\tau,t_0 )= t_0 /\pi(\tau^2+t_0^2)$ $t_0$ being the "sampling time",
and show that there exist negative lower bounds
of this weighted integrals ("quantum inequality-type relation") 
\cite{FR1,FRP,FR2}.
Yurtsver \cite{Yurts1,Yurts2} considered the condition that lower 
bound of weighted ANEC integral in certain limit is negative but 
finite (generalized ANEC). He argued that this condition 
would hold generally in four-dimensional curved spacetime, and it 
may be relevant with theorems on large scale structure implying the 
absence of macroscopic static wormholes \cite{Yurts2}.
Flanagan and Wald \cite{FW} considered contributions from the 
neighborhood of the ANEC integral by introducing the "transverse 
smearing". They show that it is enough to consider the contributions 
over several Plank lengths along the transverse direction to ensure 
the positivity of transversely smeared ANEC, which implies 
traversable wormholes have "Planck scale structures" in accordance 
with other suggestions \cite{Yurts2,FRWorm}. 
For reviewing the status of nonlocal energy 
conditions of non-interacting scalar fields, see Flanagan and 
Wald \cite{FW}.

Recently, in the extension of quantum inequality-type relation 
(quantum inequality) on flat spacetime of Ford and Roman 
\cite{FR1,FRP,FR2},
Pfenning and Ford \cite{PfF} have formulated quantum inequality 
for a static observer on a "globally static" curved spacetime 
(see section 2): The integral 
of (renormalized) energy density of a static observer with the 
weighting function $f$ has been evaluated for the minimally coupled 
scalar field in the test field limit (i.e. without considering 
the back-reaction)\footnote[4]{This limit will be assumed in this 
paper to keep the discussion manageable; However including 
back-reaction is necessary for application, as emphasized
in Ref.\cite{FW}.}, 
and showed that it has a lower bound. 
The lower bound has been given in terms of a sum of mode functions 
and evaluate explicitly for some specific spacetimes.
In the short sampling time limit where the curvature of spacetime 
could be neglected, the bound reduce to that of flat spacetime of 
Ford and Roman, to reproduce the quantum inequality in flat spacetime
\cite{FRP,FR2}. 

In this paper, we will we study the lower bound by applying the well 
established spectral theory of mathematics, to see the curvature 
effect in general curved case.
Although the spectral theory can be used more generally, in this paper
we will concentrate on the cases where the spacelike hypersurfaces of 
spacetime are compact and without boundary. 
Attentions will be focused on the short sampling time limit 
($t_0\rightarrow 0$) and the long sampling time limit 
($t_0\rightarrow \infty$). In the short sampling time limit, we 
will show that the bound has asymptotic expansion. Although the 
expansion can {\em not} be represented in general by locally 
invariant quantities (such as curvature) reflecting the nonlocal 
nature of the integral, the dominant terms in the limit turn out to be 
locally computable and are given explicitly 
in terms of the invariant quantities for 4-dimensional spacetime.
In the long sampling time limit, we will estimate how fast the bound 
goes to 0 and reveal the nonlocal nature of the bound. 
In the next section, we will briefly review the 
derivation of quantum inequality on a globally
static spacetime and introduce notations. In sections 3 and 4, the 
lower bound of quantum inequality will be studied in the short and 
long sampling time limits, respectively. The final section will be 
devoted for discussions.

\section{The quantum inequality: review}

In this section we will review the derivation of the quantum
inequality and introduce notations which will be used in this 
paper. The globally static spcetime\footnote[3]{The other name 
used widely in literature is the ultrastatic spacetime \cite{Fulling}.}
can be described by a metric of the form
\begin{equation}
ds^2= -dx_0^2 + g_{ij} ( {\bf x}) dx_i dx_j,
\end{equation}
where the $g_{ij}$ is the metric of 
the spacelike hypersurface that are 
orthogonal to the timelike Killing vector $\partial_{x_0}$. For the 
convenience of explicit evaluations, we only consider the cases 
where the spacelike hypersurfaces are compact and without boundary.
On this spacetime, the minimally coupled scalar field $\phi$ can be 
represented in terms of creation and annihilation operators
\begin{equation}
\phi= \sum_{\alpha}( a_\alpha f_\alpha + a^\dagger f_\alpha^* ),
\end{equation}
where the 
\begin{equation}
f_\alpha= e^{-i w_\alpha x_0}U_\alpha ({\bf x}) 
\end{equation}
is the positive frequency Klein-Gordon mode function. $\alpha$ is the 
set of eigenvalues which characterizes the mode functions and 
$f_\alpha$ is normalized to have unit Klein-Gordon norm. 
$U_\alpha ({\bf x})$ thus satisfy a wave equation
\begin{equation}
\nabla^i\nabla_i U_\alpha + (w_\alpha^2-\mu^2) U_\alpha =0,
\end{equation}
where $\mu$ is the constant mass and $\nabla^i $ is the covariant 
derivative operator in the three-dimensional manifold ${\cal M}$
of $t=$constant hypersurface.

The energy density to a static observer is given by
\begin{equation}
\rho= T_{\mu\nu} u^\mu u^\nu =T_{00}= \frac{1}{2}
        [(\partial_{x_0}\phi)^2 + \partial^i \phi \partial_i \phi 
          + \mu^2  \phi^2]
\end{equation}
and, making use of mode expansion form of $\phi$ in Eq.(3),
it is easy to represent the energy density in terms of creation and
annihilation operators. 
Of course, the $T_{00}$ involves the terms which diverge upon summation 
and is not well defined. We may define the renormalized $:T_{00}:$ 
through the 
normal ordering with respect to the Fock vacuum $|0>$ 
\begin{equation}
:T_{00}:= T_{00} - <0|T_{00}|0>
\end{equation}
where the vacuum would be defined by the globally timelike Killing 
vector \cite{Fulling,Kay}. 
The "averaged energy density difference" is defined by the integral 
with weighting function $f(x_0,t_0)$ as
\begin{equation}
\hat{\rho} \equiv \frac{t_0}{\pi}
   \int_{-\infty}^\infty \frac{<\psi|:T_{00}:|\psi>}{x_0^2+t_0^2} dx_0.
\end{equation}
After some algebra along the line of Ref.\cite{FRP,FR1},
$\hat{\rho}$ for arbitrary $|\psi>$ can be shown to have a lower 
bound as \cite{PfF} 
\begin{equation}
\hat{\rho} \geq -\sum_\alpha (w_\alpha^2+ 
          \frac{1}{4}\nabla^i \nabla_i )|U_\alpha |^2e^{-2w_\alpha t_0}.
\end{equation}
This quantum inequality holds on any globally static spacetime and thus 
reproduce the known inequalities for static observer. For example, 
in 4D Minkowski spacetime it gives the quantum inequality of 
Ref.\cite{FR1,FRP,FR2}. For a static observer on a circle of 
two-dimensional spacetime, it also reproduce the 
"difference inequality", Eq.(23) of Ref.\cite{FR2}.  

In this paper, we will consider only the four-dimensional spacetime. 
For the convenience of explicit evaluation, we will concentrate 
ourselves on the cases where the spacelike hypersurface ${\cal M}$ 
of the spacetime is compact and without boundary. 

If we denote the Laplacian ($-\nabla^i\nabla_i$) on ${\cal M}$ as  
$\trup$, the $w_\alpha^2$ is a egeinvalue of 
$\trup + \mu^2$  and $U_\alpha$ is the corresponding eigenfunction. 
Since the lower bound, the right hand side (r.h.s.) 
of Eq.(9), can be given 
in terms of mode functions on ${\cal M}$, instead of 
$\{ w_\alpha^2,  U_\alpha \}$ we will use the spectral resolution 
$\{ \lambda_i, \phi_i\}$ of
\begin{equation}
(\trup + \mu^2)\phi_i=\lambda_i\phi_i,\;\;\;\;\;\; 
\int_{\cal M} \phi_i^* \phi_j dvol(g)= \delta_{ij},
\end{equation}
for explicit evaluation.

The zero mode has no contribution to the bound since it can not survive 
the differentiation, and the difference in the resolutions is only in 
the normalization. Taking these points into considerations, the 
inequality can be written as   
\begin{equation}
\hat{\rho} \geq 
     -\frac{1}{2}\sum_{i=1}(\sqrt{\lambda_i} 
            -\frac{1}{4\sqrt{\lambda_i}}\trup)
          |\phi_i |^2e^{-2\sqrt{\lambda_i} t_0}.
\end{equation}
In the summation, the smallest value of $i$ is given as 1 to denote 
that zero mode case is excluded. 

By the exponential factor, it is clear that the lower bound of 
$\hat{\rho}$ converges for any positive $t_0$ and reduces to 
0 in the long sampling time limit.

\section{The short sampling time limit}
In this section we will examine the lower bound in the short sampling
time limit. For this purpose, it is useful to study the 
pseudo($\Psi$DO)-differential operator $P$ $(:=\sqrt{\trup + \mu^2})$ 
and its heat kernel. The heat kernel for $P$ in the coincident 
limit (or, on diagonal) is written as
\begin{equation} 
h({\bf x},P)(t):=\sum_i e^{-\sqrt{\lambda_i} t} |\phi_i (\bx)|^2.
\end{equation}
In the following subsection, we will evaluate the asymptotic expansion 
form of $h({\bf x},P)(t)$ as $t$ goes to 0, relying on the well-known 
facts in spectral theory of mathematics. The results will be then
applied in the subsection 3.2, to find some explicit 
form of the asymptotic expansion of the bound.

\subsection{Heat kernel method}
For the $\Psi$DO-differential operator of order 1, it has been known 
that $h({\bf x},P)(t)$ has the asymptotic expansion as $t$ goes to 0 
\cite{DG} 
\begin{equation}
h(\bx,P)(t)\sim \sum_{n=0}^\infty a_n(\bx,P) t^{n-m}
    + \sum_{k=1}^\infty b_k(\bx,P)t^k \ln t.
\end{equation}
Here the $m$ is the dimension of the manifold ${\cal M}$ on which 
the operator $P$ acts. For the 4-dimensional spacetime, $m$, the 
dimension of spacelike hypersurface, is 3. 

To determine the coefficients 
$a_n(\bx,P)$, $b_k(\bx,P)$, it is useful to define $\zeta(\bx,P,s)$
\be
\zeta(\bx,P,s):= \sum_i \lambda_i^{-s/2} |\phi_i(\bx)|^2
\ee
which is related to $h(\bx,P)(t)$ through the Mellin transformation
\be
\Gamma(s) \zeta(\bx,P,s)= \int_0^\infty t^{s-1} h(\bx,P)(t) dt.
\ee
As shown in appendix A, the coefficients can be read
from the pole structures of $\Gamma(s) \zeta(\bx,P,s)$ 
as
\begin{eqnarray}
a_n(\bx,P)&=& {\rm Res}_{s=m-n} \Gamma(s)\zeta(\bx,P,s)~~
                (={\rm residue} 
         ~{\rm of~} \Gamma(s)\zeta(\bx,P,s)~{\rm at~} s=m-n)\\
b_k(\bx,P)&=& {\rm Res}_{s=-k} (s+k)\Gamma(s)\zeta(\bx,P,s).
\end{eqnarray}
In fact, it has been known \cite{DG,APP,GG} that the asymptotic 
expansion of $h(\bx,P)(t)$
determines the pole structure of $\Gamma(s) \zeta(\bx,P,s)$ 
( conversely the pole structure of $\Gamma(s) \zeta(\bx,P,s)$ 
determines the asymptotic
expansion of $h(\bx,P)(t)$ ) to give the following equality:
\be
\Gamma(s) \zeta(\bx,P,s) =
               \sum_{n=0}^N  \frac{a_n(\bx,P)}{s+(n-m)} 
               -\sum_{k=1}^N \frac{b_k(\bx,P)}{(s+k)^2} + hol,
\ee
where $N$ is large enough integer number and $hol$ represents some 
holomorphic function of $s$ which depends on $N$ \cite{Gilkey}.
  
A remark in order is that, as explained in appendix B 
(see, also Ref.\cite{APP}), 
the $b_k(P)$ is actually  0 for even $m$ and there is no log term in 
the expansion of $h(\bx,P)(t)$. 
This implies [through the similar reasons given in the next subsection] 
that, there is {\em no} log term in the inequality of {\em odd} 
dimensional spacetime  if it is expanded in terms of short sampling 
time, as can be seen in an explicit example \cite{PfF}. 
In the following we will consider only the odd $m$ cases. 

It is useful to introduce well understood $\zeta$-function for 
the differential operator $D:=\trup + \mu^2$ which is of oder 2 
\be
\zeta(\bx,D,s):= \sum_i \lambda_i^{-s} |\phi_i(\bx)|^2.
\ee
The pole structure of $\Gamma(s) \zeta(\bx,D,s)$ is known as
\be
\Gamma(s) \zeta(\bx,D,s) =
               \sum_{n=0}^N  \frac{a_n(\bx,D)}{s+\frac{n-m}{2}} +hol. 
\ee
$a_n(\bx,D)$ has been computed 
in terms of locally invariant quantities [These quantities
defined independent of the coordinate system are written as covariant
derivatives of the curvature tensor. An example would be the scalar 
curvature or the norm of the Ricci curvature]. For compact manifold 
without boundary $a_n(\bx,D)$ is zero for odd $n$.

A simple relation of two $\zeta$-functions is available from the 
definitions;
\be 
\zeta(\bx,P,s)=\zeta(\bx,D,s/2).
\ee
which yields
\be
\Gamma(s) \zeta(\bx,P,s)= \frac{\Gamma(s)}{\Gamma(\frac{s}{2})} 
            \Gamma(\frac{s}{2})\zeta(\bx, D, \frac{s}{2})
         =2\frac{\Gamma(s)}{\Gamma(\frac{s}{2})} 
             (\sum_{n=0}\frac{a_n(\bx,D)}{s+(n-m)}+hol ).
\ee
Thus the coefficient $a_n(\bx,P)$ can be written as
\be 
a_n(\bx,P)= \Res_{s=m-n} ~[2\frac{\Gamma(s)}{\Gamma(\frac{s}{2})}
              ( \sum_{l=0} \frac{a_l(\bx,D)}{s+l-m} + hol )].
\ee
For $n\leq m$, $a_n(\bx,P)$ can be easily written in terms of 
$a_n(\bx,D)$
\begin{eqnarray}
a_n(\bx,P)&=& [\frac{2\Gamma(s)}{\Gamma(\frac{s}{2})}]_{s=m-n}~ 
                   a_n(\bx,D) ~~{\rm for} ~~m>n,\\
a_m(\bx,P)&=& a_m(\bx,D).
\end{eqnarray}
For $n>m$, due to poles in gamma function the $hol$ term can make 
contribution to the residue;
For even $n$, $ a_n(\bx,P)$ given as 
\be
a_n(\bx,P)= \Res_{s=m-n}~[2\frac{\Gamma(s)}{\Gamma(\frac{s}{2})} ~hol]
\ee
which is recently proved to be not locally computable (Theorem 1.7 of 
Ref.\cite{GG}; For the application of the general theorem of 
Gilkey and Grubb to our case, see appendix C.).  
For odd $n$, $a_n(\bx,P)$ is still locally computable, i.e. can be 
represented in terms of locally invariant quantities.

Through similar analysis, one can also find that
\be
b_k(\bx,P)=-2a_{k+m}(\bx,D)~\Res_{s=-k}~ 
\frac{\Gamma(s)}{\Gamma(\frac{s}{2})}
\ee
which is nonvanishing only for {\em odd} $k$.

Making use of the well-known coefficient $ a_n(\bx,D)$,
we can find some of the coefficients. For $m=3$ which corresponds to 
four-dimensional spacetime,
\begin{eqnarray}
a_0(\bx,P)&=& \frac{2\Gamma(s)}{\Gamma(\frac{s}{2})}\mid_{s=3} 
   ~a_0 (\bx,D)
             = \frac{1}{\pi^2}\\
a_2(\bx,P)&=&\frac{1}{4\pi^2}(\frac{\tau}{6} -\mu^2) \\
b_1(\bx,P)&=&\frac{1}{2880\pi^2} 
          (180\mu^4-60\tau \mu^2 +12\nabla_i\nabla^i\tau
              +5\tau^2-2|\rho|^2 +2|R|^2),
\end{eqnarray}
where the Ricci tensor, the scalar curvature, and the norms of the 
Ricci and full curvature tensors are defined 
as
\begin{eqnarray}
\rho_{ij}:={R_{kij}}^k &{,}& \tau:={\rho_i}^i~~{\rm and} \\
|\rho|^2:=\rho_{ij}\rho^{ij} &{,}& |R|^2:=R_{ijkl}R^{ijkl}.
\end{eqnarray}
As already explained $a_1=a_3=0$ and $a_4(\bx,P)$ is not locally 
computable.

In summary, 
\begin{eqnarray}
h(\bx,P)(t)& = &\frac{1}{\pi^2 ~t^3} +\frac{a_2(\bx,P)}{t} 
           + b_1(\bx,P)~t\ln t \nonumber \\
     && + {\rm terms~vanishing~in~ the~} t\rightarrow 0~{\rm limit}.
\end{eqnarray}
Generally, the terms vanishing in the limit $t\rightarrow 0$ could be 
locally noncomputable.
  
\subsection{ The asymptotic expansion of the lower bound }

For the evaluation of the lower bound, we need to evaluate the 
summations
\be
H_\pm
=\sum_{i=1}^\infty (\lambda_i)^{\pm \frac{1}{2}}e^{-t\sqrt{\lambda_i}}
                   |\phi_i|^2.
\ee
As in the previous subsection, these summation may be studied by 
investigating the pole structures of $\Gamma(s){\cal S}_\pm$
where 
\be
{\cal S}_\pm:=\sum_{i=1}^\infty 
           (\lambda_i)^{-s\pm\frac{1}{2}} |\phi_i|^2.
\ee

Instead, we find the asymptotic expansion form of $H_+$ 
\be
H_+= -\frac{3}{\pi^2 ~t^4} -\frac{a_2(\bx,P)}{t^2} + b_1(\bx,P)~\ln t
       + {\rm finite~or~vanishing~terms~in~ the~} 
       t\rightarrow 0{~ \rm limit}
\ee
here by differentiating the r.h.s. of Eq.(33) with respect to $t$. 
[Such operation may be justified by that the series converge in the 
$C^\infty$ topology for any finite $t$ as implied by the Lemma 1.6.5 
of Ref.\cite{Gilkey}]

For evaluation of the $H_-$, we will integrate $h(\bx,P)$;
\begin{eqnarray}
H_-&=& \int_t^\infty h(\bx,P)(y)~ dy \\
   &=& \int_t^\delta h(\bx,P)(y)~ dy +\int_\delta^\infty 
            h(\bx,P)(y)~ dy,
\end{eqnarray}
for $0<t<\delta<1$. From the fact that the heat kernel decays 
exponentially fast as $t$ goes to $\infty$, 
one can find that the second integral of the r.h.s. of Eq.(38) is 
finite.  Making use of the asymptotic expansion of $h(\bx,P)(t)$ in 
Eq.(33), from the first integral we can find the terms which diverges
in the $t\rightarrow 0$ limit
\be
H_-=\frac{1}{2\pi^2t^2}-a_2(\bx,P)\ln t + 
          {\rm finite~or~vanishing~terms~in~ the~} 
           t\rightarrow 0{~ \rm limit}.
\ee
Thus the quantum inequality of Eq.(11) can be written in the short 
sampling limit as\footnote[8]{In circulating the manuscript, I've 
been informed that the same formula has been obtained by Pfenning 
and Ford through different method \cite{new}.}
\begin{eqnarray}
\hat{\rho} &\geq& -\frac{3}{32\pi^2}[\frac{1}{t_0^4}
      +\frac{1}{3t_0^2}(\frac{\tau}{6}-\mu^2) \nonumber\\
     &&~~~ +\frac{\ln(E_0t_0)}{540} 
    (180\mu^4-60\tau \mu^2 
     -18\nabla_i\nabla^i\tau+5\tau^2-2|\rho|^2 +2|R|^2)]
       \nonumber\\
      &&~~~+{\rm finite~or~vanishing~terms~in~ the~} 
       t_0\rightarrow 0{~ \rm limit}.
\end{eqnarray}
Here, a constant $E_0$ of mass dimension has been introduced to make
the argument of logarithmic function dimensionless. 
As is clear from it's derivation, the finite or vanishing terms of 
Eq.(40) in the limit can {\em not} be represented by the locally 
invariant quantities in general.

In passing, we note that, the short sampling time expansion of
the bound for {\em odd} dimensional spacetime can be represented
by the locally invariant quantities.

For the case that ${\cal M}$ is $S^3$ of radius $a$, 
the locally invariant quantities are given as
\be
\tau=\frac{6}{a^2},~~~|\rho|^2 =|R|^2=\frac{12}{a^4}.
\ee
The quantum inequality is thus given as
\begin{eqnarray}
\hat{\rho}&\geq&-\frac{3}{32\pi^2}[\frac{1}{t_0^4}
      +\frac{1}{3t_0^2}(\frac{1}{a^2}-\mu^2) 
      +\frac{\ln(E_0t_0)}{3}(\mu^4 -2 \frac{\mu^2}{a^2}+\frac{1}{a^4})]
             \nonumber\\
    &&~~~+{\rm finite~or~vanishing~terms~in~ the~} 
      t_0\rightarrow 0{~ \rm limit},
\end{eqnarray}
in agreement with the results of Ref.\cite{PfF}.  

\section{The long sampling time limit}  
The lower bound reduces to zero in the long sampling time limit.
Even though the $T_{00}$ is normal ordered with respect to the vacuum
defined by the globally timelike killing vector, it implies 
at least that the generalized AWEC in the sense of Ref.\cite{Yurts2}
is satisfied for the timelike geodesic of static observer.

In this section, we will study how fast the lower bound will go to 
zero in the limit. This will, as a byproduct, expose the non-local 
nature of the lower bound, as already has been encountered through 
the fact that the finite or vanishing
term in the short sampling time limit can not be represented by the 
locally invariant quantities in general. 

For positive $v$ and $y~(\geq\lambda_1>0)$, the upper bound of the 
function $y^j e^{-v  \sqrt{y}}$ can be given as $C(j, v,\lambda_1)$. 
The following relation is thus true for positive $\alpha$ $(< 1)$
\begin{eqnarray}
\sum_{i=1} \lambda_i^j e^{-u\sqrt{\lambda_i}} &=& 
   \sum_{i=1}\lambda_i^j e^{-u\alpha\sqrt{\lambda_i}} 
     e^{-u(1-\alpha)\sqrt{\lambda_i}}   \nonumber\\
\leq
     C(j, u\alpha, \lambda_1) \sum_{i=1} e^{-u\sqrt{\lambda_i}}
&=&C(j, u\alpha,\lambda_1) e^{-u\sqrt{\lambda_1}(1-\alpha)}\sum_{i=1} 
   e^{-u(1-\alpha)(\sqrt{\lambda_i}-\sqrt{\lambda_1})}.
\end{eqnarray}
Furthermore since the summation 
$\sum_{i=1} e^{-u(1-\alpha)(\sqrt{\lambda_i}-\sqrt{\lambda_1})}$ is 
finite for positive $\alpha$ $(< 1)$, one can find a lemma
\be
\sum_{i=1} \lambda_i^j e^{-u\sqrt{\lambda_i}} \leq 
C'(j,d_0,\lambda_1)e^{-u\sqrt{\lambda_1}}
\ee 
in the $u\rightarrow \infty$ limit with fixed $d_0$ $(=u\alpha)$.
Here, note that the $C'(j,d_0,\lambda_1)$ do not depend on $u$.

Since $|\phi_i|^2$ or it's covariant derivative is finite, 
indeed this lemma dictates that  the lower bound reduce to 0 as fast 
as $-|f(\lambda)|e^{-2t_0\sqrt{\lambda_1}}$ in the long sampling time 
limit, where the $\lambda_1$ is the smallest one among positive 
eigenvalues of $D$.  Note that $f$ does not depend on $t_0$.
In general, the spectrum of an operator contains informations on 
global property of manifold and can not be represented only by the 
locally invariant quantities. This behavior of the bound in the long 
sampling time limit thus clearly shows the 
nonlocal nature of the bound. 

As a simple test, for the massless theory on 
${\cal M}=S^3$ (of radius $a$), our estimate yields that 
the lower bound is proportional to $-e^{-2t_0\sqrt{3}/a}$ in the 
long sampling time limit, which is in agreement with the explicit 
calculation of Ref.\cite{PfF}. 
For the massive theory, the estimate simply gives that the bound is 
proportional to $-e^{-2\mu t_0}$ in the limit for any ${\cal M}$.

\section{Discussion}
The quantum inequality or weighted AWEC exposes the distribution of
allowed negative energy density which could be neglected in simple
average along time. The weighting function thus necessarily has peak(s) 
around which main contribution to  the integral is made. 
As in the application of quantum inequality for the wormhole 
geometry via extrapolation \cite{FRWorm}, in the semiclassical 
approximation of gravity it seems unavoidable to compare the weighted 
integral with the local invariant quantities of spacetime for 
applications. Our results in previous sections show a possible nature 
of the weighted integral: it could be nonlocal.
[In fact, in the test field limit without considering back-reaction,
only the energy tensors have quantum corrections which 
contain information on global structure of spacetime.]
We think, this fact would 
play a role in finding useful weighting (sampling) functions. 
In the sense that the dominant terms of the lower bound of quantum 
inequality in the short sampling time limit are locally invariant, 
the sampling function of Ford and Roman would be still 
useful in general globally static spacetime, although applications 
of the inequality are not intended in this paper.    

The case that spacelike hypersurface is noncompact, could be physically
interesting. Although we have not treat the case in this paper, the 
heat kernel method could be useful for the evaluation of the bound
if the hypersurface is complete.

For the case that the hypersurface is compact and with boundary, the 
spacetime is not globally hyperbolic. However, the spectral theory 
itself  has been well developed and can be used in analyzing the 
bound given in  Eq.(11), as in the boundaryless case we have 
considered. In both of  the boundary conditions Neumann and Dirichlet, 
the $a_0(\bx,P)$ is same to that of Eq.(28) for four-dimensional 
spacetime, and the leading divergence of the bound in the short 
sampling time limit thus does not depend on the boundary condition. 
The sub-leading divergences, however, crucially depend on the 
boundary conditions; For example, in general $a_1(\bx,P)$ is not zero 
and depends on boundary condition \cite{Gilkey},
so that the term proportional to $1/t^3$ could appear in the short 
sampling time expansion of the bound.    

\bigskip
\begin{center}{\bf ACKNOWLEDGMENT}\end{center}
The author thanks Larry Ford, S.A. Fulling, and Tom Roman for 
discussions. It is also his pleasure to thank Peter B. Gilkey for his 
lectures as well as discussions on the spectral theory and on 
Ref.\cite{GG}, including those via J.H. Park. He acknowledges, with 
thanks, that the coefficients were actually evaluated 
by J.H. Park. This work is supported in part by nondirected research 
fund, Korea Research Foundation. 

\newpage
\section*{Appendix A}
In this appendix, we wish to expose explicitly the relations 
(Eqs.16-18) 
between the asymptotic expansion of $h(\bx,P)(t)$ and pole structure of 
$\Gamma(s)\zeta(\bx,P,s)$ \cite{DG,APP,GG,Gilkey}.
Mellin transformation of Eq.(15) gives the following equality: \\\\
$\Gamma(s)\zeta(\bx,P,s)=\int_0^1 t^{s-1} h(\bx,P)(t) dt 
                        +\int_1^\infty t^{s-1} h(\bx,P)(t) dt.
$\hfill (A1)\\
\\
Since there exist positive constants $c$ and $k$ satisfying
\[ h(\bx,P)(t) \leq ke^{-ct} ~{\rm for}~ t\geq1,\]
the integral $\int_1^\infty t^{s-1} h(\bx,P)(t) dt$ gives only 
holomorphic function of $s$. Therefore, the poles of 
$\Gamma(s)\zeta(\bx,P,s)$ should come from the first integral on 
the r.h.s. of Eq.(A1). 
Making use of the asymptotic expansion in Eq.(13),
the first integral can be calculated as \\\\
$\Gamma(s)\zeta(\bx,P,s)= \sum_{n=0}a_n\int_0^1 t^{s+n-m-1} dt
        +\sum_{k=1} b_k \int_0^1 t^{s+k-1}\ln t ~dt + hol.$
\hfill (A2)\\
\\
Reflecting the fact that $\Gamma(s)\zeta(\bx,P,s)$ has meromorphic 
extension to whole complex plane, 
except some values of $s$, the integral of 
Eq.(A2) can be carried out to give the relation of Eq.(18).

\section*{Appendix B}
In this appendix, when $m$ is even, we wish to explicitly reprove that 
there is no log term in the asymptotic expansion 
of $h(\bx,P)(t)$ \cite{APP}.
We first study $\zeta(\bx,D,s)$ at $s= -1/2,-1,-3/2,\cdots$. 
By the Eq.(20), $\zeta(\bx,D,s)$ can be written as\\\\
$\zeta(\bx,D,s) = 
   \Gamma(s)^{-1}\sum_{n=0}\frac{a_n(\bx,D)}{s+\frac{n-m}{2}}
                  +\Gamma(s)^{-1} hol.$ \hfill (B1)\\\\
Since $\Gamma(s)^{-1}$ has simple zero when $s$ is negative integer, 
Eq.(B1) shows that $\zeta(\bx,D,s)$ is regular at $s=-1,-2,\cdots$.
For $s= -1/2,-3/2,\cdots$, $\Gamma(s)^{-1}$ is regular and $m-2s$ 
is odd.  If there exist simple pole of $\zeta(\bx,D,s)$ 
for the negative half-integer $s$,
then it should be proportional to $a_{m-2s}(\bx,D)$ which is zero 
because of the odd $m-2s$. Now we have the fact that 
$\zeta(\bx,D,s)$ is regular at $s= -1/2,-1,-3/2,\cdots$.

If we assume that $b_k(\bx,P)$ is not zero, the Eqs.(17,21) then 
lead to a wrong conclusion that $\zeta(\bx,D,s/2)$ has simple 
poles for negative integer $s$, to show $b_k(\bx,P)$=0. 
$h(\bx,P)(t)$ thus does not have the log term 
in the asymptotic expansion for even $m$.

\section*{Appendix C}
In order to see how the general theorem of Gilkey and Grubb 
mentioned in the text works in our case, let's consider the scaling 
transformation
\[g_{ij}\rightarrow c^{-2}g_{ij},\;\;\; \mu^2\rightarrow c^2\mu^2\]
with positive $c$. The $P(g(c^{-2}),c^2\mu^2)$ denotes the 
scale-transformed operator from the original $P$ $(=P(g,\mu^2))$. 
The spectral resolution for $P(g(c^{-2}),c^2\mu^2)$ is then 
$\{ c\sqrt{\lambda_i}, c^{m/2}\phi_i \}$,
with the equality $P(g(c^{-2}),c^2\mu^2)=cP(g,\mu^2)$. From 
these facts, one can find the relation\\
\\
$h(\bx,P(g(c^{-2}),c^2\mu^2))(t)=c^{-m}h(\bx,P(g,\mu^2))(\frac{t}{c}) $
\hfill (C1)\\\\
which yields the equality \\\\
$a_{m+k}(\bx,P(g(c^{-2}),c^2\mu^2)) =\\
       c^{m+k} [a_{m+k}(\bx,P(g,\mu^2))+ b_k(\bx,P(g,\mu^2))\ln c ]
        ~~~{\rm for} ~~~
       k\geq 1 .$\hfill (C2)\\\\
For the odd $m$ and odd $k$ where $b_k$ is not zero, the equality of 
(C1) contradicts the Eq.(1.8) of Ref.\cite{GG} which comes from the 
assumption that $a_{m+k}$ is locally computable (See also 
Ref.\cite{Gilkey}, and note the 
slight difference in notations.). This contradiction thus shows that 
$a_{m+k}(\bx,P)$ is not locally computable for odd $m$ and $k$, 
as in the original proof of the theorem.

\end{document}